\magnification \magstep1
\raggedbottom
\openup 2\jot
\voffset6truemm
\def\cstok#1{\leavevmode\thinspace\hbox{\vrule\vtop{\vbox{\hrule\kern1pt
\hbox{\vphantom{\tt/}\thinspace{\tt#1}\thinspace}}
\kern1pt\hrule}\vrule}\thinspace}
\centerline {\bf NEW KERNELS IN QUANTUM GRAVITY}
\vskip 1cm
\leftline {Giampiero Esposito}
\vskip 0.3cm
\noindent
{\it Istituto Nazionale di Fisica Nucleare, Sezione di Napoli,
Mostra d'Oltremare Padiglione 20, 80125 Napoli, Italy}
\vskip 0.3cm
\noindent
{\it Universit\`a di Napoli Federico II, Dipartimento di Scienze
Fisiche, Complesso Universitario di Monte S. Angelo, Via Cintia,
Edificio G, 80126 Napoli, Italy}
\vskip 1cm
\noindent
{\bf Abstract.}
Recent work in the literature has proposed the use of non-local
boundary conditions in Euclidean quantum gravity. The present
paper studies first a more general form of such a 
scheme for bosonic gauge theories, by adding to
the boundary operator for mixed boundary conditions of local
nature a $2 \times 2$ matrix of pseudo-differential operators
with pseudo-homogeneous kernels. The request of invariance of
such boundary conditions under infinitesimal gauge transformations
leads to non-local boundary conditions on ghost fields. In
Euclidean quantum gravity, an alternative scheme is proposed, 
where non-local boundary conditions and the request of their
complete gauge invariance are sufficient to lead to gauge-field
and ghost operators of pseudo-differential nature. The resulting 
boundary conditions have a Dirichlet and a pseudo-differential
sector, and are pure Dirichlet for the ghost. This approach is
eventually extended to Euclidean Maxwell theory.
\vskip 100cm
\leftline {\bf 1. Introduction}
\vskip 0.3cm
\noindent
Recent efforts in the literature have led to an elegant and
useful characterization of gauge-invariant boundary
conditions for linearized quantum gravity and other fundamental
field theories. The basic elements of the Euclidean
formulation are as follows [1].

A vector bundle $V$ with connection $\nabla$ is given over an
$m$-dimensional Riemannian manifold $(M,g)$. Physical fields
are the smooth sections of $V$, and
a part of the boundary conditions is obtained by applying a
projector $\Pi$ to the field $\varphi$:
$$
[\Pi \varphi]_{\partial M}=0
\eqno (1.1)
$$
where $\partial M$ is the smooth boundary of $M$. 
Infinitesimal gauge transformations map $\varphi$ into
$$
{ }^{\varepsilon}\varphi=\varphi+R \varepsilon
\eqno (1.2)
$$
where $R$ is the generator and $\varepsilon$ is a 
`gauge function' (for gravity, $\varepsilon$ is a 1-form; for
Yang--Mills, $\varepsilon$ is a smooth function on $M$). 
Remaining boundary conditions are obtained by requiring 
that the action on $\varphi$ of the adjoint generator 
$\overline R$ should vanish on the boundary, i.e.
$$
[{\overline R}\varphi]_{\partial M}=0
\eqno (1.3)
$$
where ${\overline R} \equiv \gamma^{-1}R^{\dagger}E$, $E$
and $\gamma$ being the Hermitian metrics on the vector
bundles of physical fields and `gauge functions', respectively.
Equation (1.3) is equivalent to setting to zero on the boundary
the gauge-averaging functional (recall that Gaussian averages
over gauge functionals are necessary to obtain an invertible
gauge-field operator in the one-loop quantum theory [1]). The 
boundary conditions (1.1) and (1.3) can be re-expressed in the
convenient form [1] 
$$
\pmatrix{\Pi & 0 \cr \Lambda & I-\Pi \cr}
\pmatrix{[\varphi]_{\partial M} \cr 
[\varphi_{;N}]_{\partial M} \cr}=0
\eqno (1.4)
$$
where $\Lambda$ is a first-order differential operator on the
boundary:
$$
\Lambda \equiv (I-\Pi)\left[ 
{1\over 2} \Bigr(\Gamma^{i}{\widehat \nabla}_{i}
+{\widehat \nabla}_{i}\Gamma^{i}\Bigr)+S \right](I-\Pi)
\eqno (1.5)
$$
$\Gamma^{i}$ being a set of vector fields on $\partial M$,
$\widehat \nabla$ being the induced connection on $\partial M$,
and $S$ being an endomorphism (e.g. a term proportional to
the trace of the extrinsic-curvature tensor of the boundary).
Of course, $;N \equiv N^{a}\nabla_{a}$ 
in Eq. (1.4) is the standard notation for the
covariant derivatives along the inward-pointing normal to the
boundary. 

Boundary conditions of the form (1.4) were first considered, 
from the point of view of pure mathematics, by Gilkey and 
Smith [2], and have been the object of increasing interest
in the analysis of heat-kernel asymptotics [3--7], 
string theory [8, 9] and Euclidean quantum
gravity [10--14]. On the other hand, it is precisely the attempt 
of applying such a framework to the one-loop semiclassical
approximation for the quantized gravitational field which
has led to the discovery of a serious technical problem,
i.e. the impossibility of achieving a strongly elliptic 
boundary value problem [1, 15]. This is a mathematical condition
on the uniqueness of the solution of the eigenvalue equation
for the leading symbol $\sigma_{L}(P)$ of the gauge-field 
operator $P$ when restricted to the boundary [1], 
subject to a decay condition at infinite geodesic
distance from the boundary and to the boundary conditions of
the problem. If strong ellipticity fails to hold, the leading
symbol $\sigma_{L}(P)$ is no longer invertible, and the functional
trace of the heat operator acquires a non-integrable part as one
approaches the boundary along the inward geodesic flow. The 
$L^{2}$-trace of ${\rm e}^{-tP}$ has then an asymptotic expansion,
as $t \rightarrow 0^{+}$, which contains not only the standard
interior and boundary terms [16], but also non-integrable 
contributions [1]. This makes it impossible to use heat-kernel
methods (only) to interpret the ultraviolet divergences of the
corresponding one-loop quantum theory.

In [17] it has been therefore suggested to look at boundary
conditions in Euclidean quantum gravity from the point of view
of pseudo-differential boundary value problems. Indeed, not only
has the functional calculus of pseudo-differential boundary 
problems [18] reached a status of high rigour, with several 
interesting results of general nature, but relevant physical 
applications are already available in lower-dimensional 
theories [19]. In the case of gravitational
perturbations, the work in [17] has added to the local boundary
operator in Eq. (1.4) a non-local boundary operator with kernel
$W_{ab}^{\; \; \; cd}(x,x')$, and special assumptions on the
components of $W$ have been made to simplify the resulting
calculations. Here we follow instead a more general path
because we would like it to understand how far can one go
in allowing for non-local boundary data. For this purpose,
section $2$ considers non-local boundary 
conditions for any bosonic gauge-field theory. Section 3 
studies the invariance properties of the boundary operator 
with the associated consistency conditions. 
An alternative set of non-local boundary conditions
in Euclidean quantum gravity are proposed and studied in section 4,
and their counterpart in Euclidean Maxwell theory is derived
in section 5. Concluding remarks are presented
in section 6, and relevant details are described in the appendix.
\vskip 0.3cm
\leftline {\bf 2. A new set of non-local boundary conditions} 
\vskip 0.3cm
\noindent
For bosonic field theories, non-local boundary conditions add an
integral operator to the projectors, endomorphisms and 
first-order differential operators occurring in the purely local 
theory (see Eq. (1.4)). For example, as we already said in [17]
following [18], in population theory one studies the equation
$$
u(0)-\int_{0}^{\infty}u(t)f(t)dt=0.
\eqno (2.1)
$$
We here consider a pseudo-differential boundary operator 
defined by a kernel [16], so that Eq. (1.4) is replaced
by the boundary condition
$$ 
\pmatrix{\Pi + \alpha & \beta \cr 
\Lambda + \gamma & I-\Pi+\delta \cr}
\pmatrix{[\varphi]_{\partial M} \cr 
[\varphi_{;N}]_{\partial M} \cr}=0.
\eqno (2.2)
$$
In Eq. (2.2) $\alpha, \beta, \gamma$ and $\delta$ should be viewed 
as integral operators with kernels $A,B,C$ and $D$,
respectively. In the case of metric perturbations $h_{ab}$, 
their action is taken to be
$$
(\alpha h)_{ab}(x)
\equiv \int_{M}A_{ab}^{\; \; \; pq}(x,x-x')h_{pq}(x')dV'
\eqno (2.3)
$$
$$
(\beta h)_{ab}(x)
\equiv \int_{M}B_{ab}^{\; \; \; pq}(x,x-x')h_{pq}(x')dV'
\eqno (2.4)
$$
$$
(\gamma h)_{ab}(x)
\equiv \int_{M}C_{ab}^{\; \; \; pq}(x,x-x')h_{pq}(x')dV'
\eqno (2.5)
$$
$$
(\delta h)_{ab}(x)
\equiv \int_{M}D_{ab}^{\; \; \; pq}(x,x-x')h_{pq}(x')dV'
\eqno (2.6)
$$
where $dV'$ is the integration measure over $M$, i.e.
$$
dV' \equiv \sqrt{{\rm det} \; g(x')} \; dx'^{1}...dx'^{m}.
$$
The form of the integrands in (2.3)--(2.6) is consistent with our
assumption of dealing with a pseudo-differential boundary operator.
This is a technical point which is described in detail in the appendix.
With the notation defined therein, we say that $A,B,C$ and $D$ are
taken to be {\it pseudo-homogeneous kernels}. Note that Eq. (2.2) requires
taking the restriction to the boundary of what is obtained from
(2.3)--(2.6), e.g. (cf [17])
$$
\Bigr[(\alpha h)_{ab}(x)\Bigr]_{\partial M}
=\left[ \int_{M}A_{ab}^{\; \; \; pq}(x,x-x')h_{pq}(x')dV'
\right]_{\partial M}.
\eqno (2.7)
$$
\vskip 0.3cm
\leftline {\bf 3. Ghost boundary operator and consistency conditions}
\vskip 0.3cm
\noindent
Suppose that our bosonic gauge fields are subject to the
infinitesimal gauge transformations (1.2), where the form of the
generators $R$ remains, for the time being, not specified. If
both $\varphi$ and ${ }^{\varepsilon}\varphi$ satisfy the
boundary conditions (2.2), we say that such boundary conditions
are preserved under the action of $R$. The resulting boundary
conditions on gauge functions are 
$$
\Bigr[(\Pi+\alpha)R\varepsilon +\beta(R \varepsilon)_{;N}
\Bigr]_{\partial M}=0
\eqno (3.1)
$$
$$
\Bigr[(\Lambda+\gamma)R \varepsilon
+(I-\Pi+\delta)(R \varepsilon)_{;N}\Bigr]_{\partial M}=0.
\eqno (3.2)
$$
Equations (3.1) and (3.2) should provide boundary conditions on
the ghost while avoiding an over-determined boundary value problem.
For this purpose, assuming that $\Pi+\alpha$ has an inverse, we
derive from (3.1) that
$$
[R \varepsilon]_{\partial M}=-\Bigr[(\Pi+\alpha)^{-1}
\beta (R \varepsilon)_{;N}\Bigr]_{\partial M}.
\eqno (3.3)
$$
The insertion of such a formula into Eq. (3.2) leads to
$$
\biggr[\Bigr(I-\Pi+\delta -(\Lambda+\gamma)
(\Pi+\alpha)^{-1}\beta \Bigr)(R \varepsilon)_{;N}
\biggr]_{\partial M}=0.
\eqno (3.4)
$$
The condition (3.4) should hold for all $\varepsilon$, to avoid
having too many boundary conditions after having imposed Eq.
(3.3). Hence it becomes an operator equation which restricts the
form of the boundary operator, i.e.
$$
I-\Pi+\delta -(\Lambda+\gamma)(\Pi+\alpha)^{-1}\beta=0.
\eqno (3.5)
$$
It should be stressed that this equation can only make sense with
non-local boundary conditions, since otherwise $\Pi+\alpha$ would
reduce to $\Pi$, i.e. a projector for which no inverse exists
(the only map satisfying the properties $T^{2}=T$ and 
$TT^{-1}=T^{-1}T=I$ is the identity $I$).     

By contrast, if $\beta$ is invertible while $\Pi+\alpha$ cannot be
inverted, Eq. (3.1) can be cast in the form
$$
\Bigr[(R \varepsilon)_{;N}\Bigr]_{\partial M}=
-\Bigr[\beta^{-1}(\Pi+\alpha)R \varepsilon \Bigr]_{\partial M}.
\eqno (3.6)
$$
This implies, upon insertion into Eq. (3.2), the consistency 
condition
$$
\biggr[\Bigr((\Lambda+\gamma)-(I-\Pi+\delta)\beta^{-1}
(\Pi+\alpha)\Bigr)(R \varepsilon)\biggr]_{\partial M}=0.
\eqno (3.7)
$$
Since Eq. (3.7) should hold for all $\varepsilon$, it leads to
an operator equation that restricts the admissible form of the
boundary operator, i.e. (cf (3.5))
$$
\Lambda+\gamma-(I-\Pi+\delta)\beta^{-1}(\Pi+\alpha)=0.
\eqno (3.8)
$$
For example, the form (3.6) of the ghost boundary conditions
should be used when both $\Pi$ and $\Pi+\alpha$ are projection
operators, so that the condition
$$
(\Pi+\alpha)^{2}=\Pi+\alpha
\eqno (3.9)
$$
implies that
$$
\alpha^{2}+\Pi \alpha + \alpha \Pi=\alpha .
\eqno (3.10)
$$

The present analysis of consistency conditions for ghost boundary
conditions should be supplemented by the general equation which
ensures that non-trivial solutions of the system (3.1) and (3.2)
exist. This is the operator equation
$$
(\Pi+\alpha)(I-\Pi+\delta)-\beta(\Lambda+\gamma)=0.
\eqno (3.11)
$$
For example, when $\alpha, \beta, \gamma$ and $\delta$ are set
to zero, one recovers the identity $\Pi(I-\Pi)=0$ which is 
satisfied by the boundary operator for local boundary conditions
in Eq. (1.4), since $\Pi$ is a projector. 
Equation (3.11) leads to
$$
\alpha(I+\delta)-\beta \gamma + \Pi \delta 
-\alpha \Pi -\beta \Lambda=0.
\eqno (3.12)
$$
If we now use the consistency condition (3.5), we can express
$I+\delta$ in the form
$$
I+\delta=\Pi+(\Lambda+\gamma)(\Pi+\alpha)^{-1}\beta
\eqno (3.13)
$$
and its insertion into Eq. (3.12) yields
$$
\alpha (\Lambda+\gamma)(\Pi+\alpha)^{-1}\beta
-\beta(\Lambda+\gamma)+\Pi \delta=0.
\eqno (3.14)
$$

On the other hand, if we use Eq. (3.12) and the consistency
condition (3.8), we find from (3.12) that
$$
\Lambda+\gamma=\beta^{-1} \Bigr[\alpha(I+\delta)
+\Pi \delta -\alpha \Pi \Bigr]
\eqno (3.15)
$$
whilst Eq. (3.8) yields
$$
\Lambda+\gamma=(I-\Pi+\delta)\beta^{-1}(\Pi+\alpha).
\eqno (3.16)
$$
The right-hand sides of (3.15) and (3.16) should therefore
coincide, which implies (upon application of $\beta$ from the
left to both of them)
$$
\alpha(I+\delta)+\Pi \delta -\alpha \Pi
=\beta(I-\Pi+\delta)\beta^{-1}(\Pi+\alpha).
\eqno (3.17)
$$
\vskip 0.3cm
\leftline {\bf 4. Euclidean quantum gravity}
\vskip 0.3cm
\noindent
In Euclidean quantum gravity, the request of invariance under
infinitesimal diffeomorphisms of homogeneous Dirichlet conditions
on spatial components of metric perturbations leads to homogeneous
Dirichlet conditions on the whole ghost 1-form for all boundaries
which are not totally geodesic [12] (a boundary is said to be
totally geodesic when its extrinsic curvature tensor
vanishes). At that stage, the vanishing 
of the gauge-averaging functional $\Phi_{a}(h)$ at the boundary
is imposed to ensure that this remaining set of boundary conditions
on metric perturbations leads again to homogeneous Dirichlet
conditions on the ghost (with the exception of zero-modes) 
[1, 10--15]. The difference between the present paper and the 
scheme considered in [1, 10--15] lies, however, in the use of non-local
boundary operators (see also [20]). 

We therefore assume that $\Phi_{a}(h)$ consists of the de Donder
term (which has the advantage of leading to an operator of Laplace
type on metric perturbations with purely local boundary conditions)
plus a pseudo-differential part with pseudo-homogeneous kernel, i.e.
(here ${\hat h} \equiv g^{cd}h_{cd}$)
$$
\Phi_{a}(h) \equiv \nabla^{b} 
\left(h_{ab}-{1\over 2}g_{ab}{\hat h} \right)
+(\zeta h)_{a}
\eqno (4.1)
$$
where 
$$
(\zeta h)_{a}(x) \equiv \int_{M} \zeta_{a}^{\; cd}(x,x-x')
h_{cd}(x')dV'.
\eqno (4.2)
$$
As in [1], $\nabla$ is the connection on the bundle of symmetric
rank-2 tensor fields over $M$, $g$ is the background metric,
and $dV'$ is the integration measure over $M$ 
used already in (2.3)--(2.6). Moreover,
$\zeta_{a}^{\; cd}(x,x-x')$ is taken to be a pseudo-homogeneous
kernel on the curved $m$-dimensional Riemannian background
$(M,g)$. By virtue of the assumption (4.2), the standard rule for
the evaluation of the ghost operator ${\cal F}_{a}^{\; b}$ yields
now a non-trivial result, because (see (1.2))
$$
\Phi_{a}(h)-\Phi_{a}({ }^{\varepsilon}h)
={\cal F}_{a}^{\; b} \; \varepsilon_{b}
\eqno (4.3)
$$
where (here $\cstok{\ } \equiv g^{ab}\nabla_{a}\nabla_{b}$)
$$
{\cal F}_{a}^{\; b} \equiv -\delta_{a}^{\; b} \cstok{\ }
-R_{a}^{\; b}+{\widetilde {\cal F}}_{a}^{\; b}
\eqno (4.4)
$$
${\widetilde {\cal F}}_{a}^{\; b}$ being the pseudo-differential
operator defined by
$$
{\widetilde {\cal F}}_{a}^{\; b} \; \varepsilon_{b}(x)
\equiv - \int_{M} \zeta_{a}^{\; cd}(x,x-x')\nabla_{(c} \; 
\varepsilon_{d)}(x')dV'.
\eqno (4.5)
$$

The gauge-field operator $P_{ab}^{\; \; \; cd}$ on metric 
perturbations is obtained by expanding to quadratic order in
$h_{ab}$ the Euclidean Einstein--Hilbert action and adding the 
integral over $M$ of ${\Phi_{a}(h)\Phi^{a}(h)\over 2\alpha}$.
On setting $\alpha=1$ for simplicity, one finds that
$$
P_{ab}^{\; \; \; cd}=G_{ab}^{\; \; \; cd}
+U_{ab}^{\; \; \; cd}+V_{ab}^{\; \; \; cd}.
\eqno (4.6)
$$
With our notation, $G_{ab}^{\; \; \; cd}$ is the operator of
Laplace type given by [13]
$$ \eqalignno{
G_{ab}^{\; \; \; cd} & \equiv E_{ab}^{\; \; \; cd}
\left(-\cstok{\ }+R \right)
-2E_{ab}^{\; \; \; qf} \; R_{\; qpf}^{c} \; g^{dp} \cr
&-E_{ab}^{\; \; \; pd} \; R_{p}^{\; c}
-E_{ab}^{\; \; \; cp} \; R_{p}^{\; d} 
&(4.7)\cr}
$$
where $E^{abcd}$ is the DeWitt supermetric (i.e. the metric on
the vector bundle of symmetric rank-2 tensor fields over $M$)
$$
E^{abcd} \equiv {1\over 2}\Bigr(g^{ac}g^{bd}+g^{ad}g^{bc}
-g^{ab}g^{cd}\Bigr).
\eqno (4.8)
$$
Moreover, $U_{ab}^{\; \; \; cd}$ and $V_{ab}^{\; \; \; cd}$ are
pseudo-differential operators resulting from
$\Phi_{a}(h)(\zeta h)^{a}$ and ${(\zeta h)_{a}(\zeta h)^{a}\over 2}$
respectively, in the expression of the gauge-averaging term 
${\Phi_{a}(h)\Phi^{a}(h)\over 2}$. To work out the former, we
use the identity
$$ \eqalignno{
\; & \nabla^{b}\left[\left(h_{ab}-{1\over 2}g_{ab}{\hat h} \right)
\int_{M}\zeta^{acd}(x,x-x')h_{cd}(x')dV' \right] \cr
& \equiv \nabla^{b}T_{b} \cr
&=\left[\nabla^{b}\left(h_{ab}-{1\over 2}g_{ab}{\hat h} \right)\right]
\int_{M}\zeta^{acd}(x,x-x')h_{cd}(x')dV' \cr
&+\left(h_{ab}-{1\over 2}g_{ab}{\hat h} \right)\nabla^{b}
\int_{M}\zeta^{acd}(x,x-x')h_{cd}(x')dV'.
&(4.9)\cr}
$$
We therefore define
$$
T_{b} \equiv \left(h_{ab}-{1\over 2}g_{ab}{\hat h}\right)
(\zeta h)^{a}
\eqno (4.10)
$$
and add to the action a boundary term equal to
($d\Sigma'$ being the integration measure over $\partial M$)
$$
-\int_{\partial M}N^{b}T_{b}d\Sigma'
$$
to find that the action of $U_{ab}^{\; \; \; cd}$ is defined by
$$
U_{ab}^{\; \; \; cd}h_{cd}(x) \equiv -2 \nabla^{r}E_{rsab}
\int_{M} \zeta^{scd}(x,x-x')h_{cd}(x')dV'.
\eqno (4.11)
$$
Furthermore, the definition (4.2) implies that 
$V_{ab}^{\; \; \; cd}$ is a pseudo-differential operator whose 
action is defined by
$$ \eqalignno{
\; & h^{ab} \; V_{ab}^{\; \; \; cd} \; h_{cd}(x) \cr
& \equiv \int_{M^{2}} h^{ab}(x') \zeta_{pab}(x,x-x')
\zeta^{pcd}(x,x-x'')h_{cd}(x'')dV' dV''.
&(4.12)\cr}
$$

It should be stressed that the scheme considered in the
present section differs substantially from the analysis of
sections 2 and 3, because we have imposed the following boundary
conditions on metric perturbations:
$$
[h_{ij}]_{\partial M}=0
\eqno (4.13)
$$
$$
\Bigr[\Phi_{a}(h)\Bigr]_{\partial M}
=\left[\nabla^{b}\left(h_{ab}-{1\over 2}g_{ab}{\hat h} \right)
+(\zeta h)_{a}\right]_{\partial M}=0.
\eqno (4.14)
$$  
Both (4.13) and (4.14) are invariant under infinitesimal 
diffeomorphisms if the whole ghost 1-form vanishes
at the boundary, i.e.
$$
[\varepsilon_{a}]_{\partial M}=0.
\eqno (4.15)
$$
Thus, we have proposed the general equations for an approach
to Euclidean quantum gravity where both metric perturbations and
ghost fields are ruled by pseudo-differential operators, while 
the boundary conditions have a Dirichlet and a pseudo-differential
sector (and are pure Dirichlet for the ghost). As we know from the
general analysis in [17] and [18], the adjoint of the ghost 
operator will instead be a differential operator, but subject to
non-local boundary conditions, to ensure self-adjointness.
\vskip 0.3cm
\leftline {\bf 5. Quantized Maxwell theory}
\vskip 0.3cm
\noindent
It may be now helpful to show how the scheme of section 4 can be
applied to a simpler field theory, i.e. Euclidean Maxwell theory.
For this purpose, we consider a gauge-averaging functional 
$\Phi(A)$ given by the Lorenz term (i.e. the counterpart of the
de Donder term for gravity) plus a pseudo-differential 
contribution $Q(A)$ with pseudo-homogeneous kernel 
$Q^{b}(x,x-x')$, i.e.
$$
\Phi(A) \equiv \nabla^{b}A_{b}+Q(A)
\eqno (5.1)
$$
where
$$
Q(A)(x) \equiv \int_{M}Q^{b}(x,x-x')A_{b}(x')dV'.
\eqno (5.2)
$$
With our notation, $\nabla$ is now the connection on the vector
bundle $W$ of 1-forms $A_{b}dx^{b}$ over $M$. The metric on $W$ 
coincides with the metric $g$ of $M$, and infinitesimal gauge
transformations read
$$
{ }^{\varepsilon}A_{b} \equiv A_{b}+\nabla_{b}\varepsilon.
\eqno (5.3)
$$
By virtue of the hypothesis (5.1), the ghost operator $\cal F$
is found from the equation
$$
\Phi(A)-\Phi({ }^{\varepsilon}A)={\cal F} \varepsilon
\eqno (5.4)
$$
where
$$
{\cal F} \equiv -\cstok{\ }+{\widetilde {\cal F}}
\eqno (5.5)
$$
${\widetilde {\cal F}}$ being the pseudo-differential operator
defined by
$$
{\widetilde {\cal F}}\varepsilon(x) \equiv -\int_{M}
Q^{b}(x,x-x')(\nabla_{b}\varepsilon)(x')dV'.
\eqno (5.6)
$$

The gauge-field operator $P_{a}^{\; b}$ on perturbations of the
potential (the background value of $A$ is here taken to vanish) 
is obtained by expanding to quadratic order in $A$ the Euclidean
Maxwell action and adding the integral over $M$ of
${[\Phi(A)]^{2}\over 2\alpha}$. On setting again $\alpha=1$ 
for simplicity as in section 4, one finds that
$$
P_{a}^{\; b}=G_{a}^{\; b}+U_{a}^{\; b}+V_{a}^{\; b}.
\eqno (5.7)
$$
We denote by $G_{a}^{\; b}$ the operator of Laplace type
$$
G_{a}^{\; b} \equiv -\delta_{a}^{\; b}\cstok{\ }
+R_{a}^{\; b}.
\eqno (5.8)
$$
Furthermore, $U_{a}^{\; b}$ and $V_{a}^{\; b}$ are 
pseudo-differential operators. The former is obtained by using
the identity
$$
\nabla^{b}(A_{b}Q(A))=(\nabla^{b}A_{b})Q(A)
+A_{b}\nabla^{b}Q(A).
\eqno (5.9)
$$
Thus, on defining
$$
{\widetilde T}_{b} \equiv A_{b}Q(A)
\eqno (5.10)
$$
we add to the action integral for the 1-loop quantum theory a
boundary term equal to
$$
-\int_{\partial M}N^{b}{\widetilde T}_{b}d\Sigma'
$$
to find that the action of $U_{a}^{\; b}$ is defined by
$$
U_{a}^{\; b}A_{b}(x) \equiv -2\nabla_{a} \int_{M}
Q^{b}(x,x-x')A_{b}(x')dV'.
\eqno (5.11)
$$
Last, the operator $V_{a}^{\; b}$ results from $Q^{2}(A)$ in
$[\Phi(A)]^{2}$, and is such that
$$ \eqalignno{
\; & A^{a}V_{a}^{\; b}A_{b}(x) \cr
& \equiv \int_{M^{2}} A^{a}(x')
Q_{a}(x,x-x')Q^{b}(x,x-x'')A_{b}(x'')dV'dV''.
&(5.12)\cr}
$$

The gauge-invariant boundary conditions which represent the
electromagnetic counterpart of Eqs. (4.13) and (4.14) for
gravity are (cf [13])
$$
\Bigr[(\delta_{a}^{\; b}-N_{a}N^{b})A_{b}\Bigr]_{\partial M}=0
\eqno (5.13)
$$
$$
[\Phi(A)]_{\partial M}
=\Bigr[\nabla^{b}A_{b}+Q(A)\Bigr]_{\partial M}
\eqno (5.14)
$$
where the square bracket on the left-hand side of Eq. (5.13) 
represents the tangential components of the potential.
Both Eq. (5.13) and (5.14) are preserved under the action of gauge
transformations (5.3) provided that the ghost field vanishes
at the boundary:
$$
[\varepsilon]_{\partial M}=0
\eqno (5.15)
$$
where Eq. (5.15) should also hold for the anti-ghost [13, 22].
\vskip 0.3cm
\leftline {\bf 6. Concluding remarks}
\vskip 0.3cm
\noindent
The theory of elliptic operators on compact Riemannian manifolds
$(M,g)$ with or without boundary is a fascinating topic which has led to
several deep developments in mathematics and theoretical physics [21].
There is, first, the Dirac operator obtained by composing Clifford
multiplication with covariant differentiation [21], i.e.
$$
{\cal D} \equiv \gamma^{a}\nabla_{a}
\eqno (6.1)
$$
whose symbol generates all elliptic symbols on a compact Riemannian
manifold, so that this operator may be viewed as the most fundamental
among all elliptic operators (and also very well suited to studying
topological properties of vector bundles over $M$). 

The Laplace operator or, more generally, operators of Laplace type
$$
P: C^{\infty}(V,M) \rightarrow C^{\infty}(V,M)
\eqno (6.2a)
$$
$$
P \equiv -g^{ab}\nabla_{a}\nabla_{b}-E
\eqno (6.2b)
$$
occurs whenever one tries to quantize bosonic gauge theories in
linear covariant gauges via path-integral methods [22]. Its leading 
symbol is scalar [1, 16]. 

The next step is the consideration of non-minimal operators,
whose leading symbol is not of scalar type [16]. They are
relevant for the quantization of bosonic gauge theories in 
arbitrary gauges. An example is provided by the gauge-field
operator when the Lorenz gauge-averaging term is used 
for Euclidean Maxwell theory with arbitrary gauge parameter
$\alpha$, i.e.
$$
P_{a}^{\; b} \equiv -\delta_{a}^{\; b}\cstok{\ }
+\left(1-{1\over \alpha}\right)\nabla_{a}\nabla^{b}
+R_{a}^{\; b}.
\eqno (6.3)
$$

A further step is given by the analysis of conformally covariant
operators $Q$. Their consideration is suggested by the behaviour
under conformal rescalings
$$
g_{ab} \rightarrow {\rm e}^{2\Omega} g_{ab}
\eqno (6.4)
$$
of the metric, since they are then found to transform according 
to the rule
$$
Q(\Omega)={\rm e}^{-(m+4)\Omega /2} \; Q(\Omega=0) \;
{\rm e}^{(m-4)\Omega /2}.
\eqno (6.5)
$$
In other words, conformally covariant operators occur whenever 
conformal symmetries play a key role in the process of understanding
some features of a field theory (e.g. supplementary conditions
which are conformally invariant [23, 24]).

Still, all the above operators are only particular cases of the
general family of pseudo-differential operators
[16, 18, 25]. While differential
operators $D$ are local in that the equation $f=0$ implies $Df=0$
as well, this is no longer true for pseudo-differential operators,
because their construction in ${\bf R}^{m}$ involves the Fourier
transform, which smears out the support [16]. Moreover, their 
definition can be extended to Riemannian manifolds in a
coordinate-free way [16]. A careful analysis of the role of
pseudo-differential operators in Euclidean quantum gravity is
therefore quite important, because all what has been achieved so
far [13, 26] is only a particular case of a more general framework 
which is, to a large extent, unexplored in the literature on
quantum gravity [17].

In our paper, we have first considered non-local boundary conditions in
the form (2.2), with the action of $\alpha, \beta, \gamma$ and
$\delta$ defined by polyhomogeneous kernels as in (2.3)--(2.6).
The preservation of such boundary conditions under infinitesimal
gauge transformations leads to the ghost boundary conditions (3.1)
and (3.2). Actually, Eq. (3.1) has been required to specify completely
the behaviour of ghost fields at the boundary. Equation (3.2) 
becomes then a consistency condition which restricts the admissible
forms of non-local boundary operators (see (3.5) and (3.8)). 
On the other hand, non-trivial solutions of the system given by
(3.1) and (3.2) exist if and only if Eq. (3.11) holds. Its
compatibility with Eq. (3.5) leads to Eq. (3.14), whereas its
compatibility with Eq. (3.8) yields Eq. (3.17).

In Euclidean quantum gravity, an alternative scheme has been built,
with gauge-averaging functional given by (4.1) and (4.2), ghost
operator in the form (4.4) and (4.5), operator on metric perturbations
described by (4.6), (4.7), (4.11) and (4.12). Interestingly, 
non-local boundary conditions, jointly with the request of their
complete gauge invariance, are sufficient to lead to gauge-field
and ghost operators of pseudo-differential nature. The
resulting ghost boundary conditions are homogeneous Dirichlet as
in (4.15), and hence differ from (3.1) and (3.2).

Section 5 has shown in detail how the formalism of section 4 can 
also be applied to the 1-loop semiclassical theory of the free
Maxwell field in vacuum. Once more, gauge-invariant boundary
conditions on the gauge field given by a direct sum of a
Dirichlet and a pseudo-differential sector are sufficient to
lead to gauge-field and ghost operators of pseudo-differential
nature.

It now remains to be seen which one, among the schemes of sections
3 and 4, is best suited to satisfy the requirement of strong
ellipticity of the pseudo-differential boundary value problem
(see definitions and theorems in section 1.7 of [18]). Moreover,
kernels of distributional nature, rather than the polyhomogeneous
kernels advocated in our paper, might be studied as well, 
generalizing to curved backgrounds the analysis in [27]. At the
present stage, it is unclear whether both families of kernels are
needed in quantum gravity, or whether one of the two is more
fundamental. Last, but not least, it appears desirable to investigate
the role of non-local boundary conditions 
in quantum cosmology [13, 26, 28],
with application to 1-loop semiclassical effects and 
large-scale structure of the universe.
\vskip 0.3cm
\leftline {\bf Acknowledgments}
\vskip 0.3cm
\noindent
This work has been partially supported by PRIN97 `Sintesi'.
The author is much indebted to Ivan Avramidi and Alexander 
Kamenshchik for years of joint work on the boundary conditions
in Euclidean quantum gravity. He would like to dedicate the
present paper to Michela Foa.
\vskip 0.3cm
\leftline {\bf Appendix}
\vskip 0.3cm
\noindent
In [17] we have given some basic definitions about pseudo-differential
operators and their symbols. In the present paper, however, we are
interested in kernels of pseudo-differential operators, and hence 
we need a more advanced treatment. For this purpose, we follow chapter
III of [25]. The basic concepts and results we need are as follows.
\vskip 0.3cm
\noindent
{\it Definition 1.} Let $\omega$ be a complex number such that 
$\omega \not = 0,1,2,...$, and ${\rm Re}(\omega) > -\nu$. A
function $f$ is a $C^{\infty}$ {\it pseudo-homogeneous function}
of degree $\omega$, denoted by $\psi h f_{\omega}$, if and only if
outside the origin $f$ is a $C^{\infty}$ function homogeneous
of degree $\omega$:
$$
f(tx)=t^{\omega}f(x).
\eqno (A.1)
$$
Moreover, if $\omega=0,1,2,...$, then $f$ is a $\psi h f_{\omega}$
if and only if
$$
f(x)=P(x)\log|x|+g(x)
\eqno (A.2)
$$
where $P$ is a polynomial of degree $\omega$, and $g$ is a 
$C^{\infty}$ function homogeneous of degree $\omega$, for which
$$
\int_{|x|=1}x^{\alpha}g(x)=0 \; \; \; 
|\alpha|=\omega.
\eqno (A.3)
$$
\vskip 0.3cm
\noindent
{\it Definition 2.} Let $U \subset {\bf R}^{\nu}$ be open, and let
$K(x,x-y)$ be defined for $x,y$ in $U$, with $x \not = y$. Let
${\rm Re}(\omega) > -\nu$. Then $K$ is a {\it pseudo-homogeneous
kernel} of degree $\omega$, denoted by $\psi h k_{\omega}$, if and
only if there exist functions $K_{\omega+j}(x,z)$, of class 
$C^{\infty}$ for $z \not = 0$ and pseudo-homogeneous in $z$ of 
degree $\omega+j$, such that
$$
K(x,x-y)-\sum_{j < J}K_{\omega +j}(x,x-y)
$$
is of class $C^{k}$ for $k < {\rm Re}(\omega)+J$.
\vskip 0.3cm
\noindent
{\it Theorem 1.} Let ${\rm Re}(\omega)<0$. Then
$$
A: C_{c}^{\infty}(U) \rightarrow C^{\infty}(U)
$$
is a pseudo-differential operator of degree $\omega$ if and 
only if 
$$
Af(x)=\int_{U}K(x,x-y)f(y)dy
\eqno (A.4)
$$
where $K$ is a pseudo-homogeneous kernel of degree 
$-\omega-\nu$.

A theory with ${\rm Re}(\omega) \geq 0$ might also be developed, 
leading to distributional kernels, as is shown in [27]. Two
general principles underlie Theorem 1:
\vskip 0.3cm
\noindent
(i) If the kernel $K$ is homogeneous of degree $s$, then its 
Fourier transform $\hat K$ is homogeneous of degree $-s-\nu$.
One can indeed write
$$
{\hat K}(t \xi)=\int {\rm e}^{ixt \xi}K(x)dx
\eqno (A.5)
$$
and upon setting $xt=y$ one finds
$$
{\hat K}(t\xi)=\int {\rm e}^{iy \xi}K(yt^{-1})t^{-\nu}dy
=t^{-s-\nu}{\hat K}(\xi)
\eqno (A.6)
$$
which proves our statement. However, the integral (A.5) is not
absolutely convergent for any value of $s$, and not even 
conditionally convergent for most values of $s$.
\vskip 0.3cm
\noindent
(ii) The behaviour of $\hat K$ near $\infty$ is closely related
to the behaviour of $K$ near $0$. This property can be exploited 
to circumvent the convergence problem.
\vskip 0.3cm
\noindent
Interestingly, one can pass from a pseudo-homogeneous kernel
$K_{s}$ to a symbol $a_{-s-\nu}$; one recovers $a_{-s-\nu}$
indirectly from the function $\xi^{\alpha}a_{-s-\nu}$ for
sufficiently large $\alpha$.

All the above definitions and properties hold for 
pseudo-differential operators on ${\bf R}^{m}$. On extending the
formalism to generic $m$-dimensional Riemannian manifolds $(M,g)$,
it remains to be proved in a rigorous way that it is sufficient
to replace $dy \equiv dy^{1}...dy^{m}$ in (A.4) by the integration
measure $\sqrt{{\rm det} \; g(y)} \; dy^{1}...dy^{m}$ (see
section 1.3 of [16]).
\vskip 0.3cm
\leftline {\bf References}
\vskip 0.3cm
\noindent
\item {[1]}
Avramidi I G and Esposito G 1999 {\it Commun. Math. Phys.}
{\bf 200} 495
\item {[2]}
Gilkey P B and Smith L 1983 {\it J. Diff. Geom.} 
{\bf 18} 393
\item {[3]}
McAvity D M and Osborn H 1991 {\it Class. Quantum Grav.}
{\bf 8} 1445
\item {[4]}
Dowker J S and Kirsten K 1997 {\it Class. Quantum Grav.} 
{\bf 14} L169
\item {[5]}
Avramidi I G and Esposito G 1998 {\it Class. Quantum Grav.}
{\bf 15} 281
\item {[6]}
Elizalde E and Vassilevich D V 1999 {\it Class. Quantum Grav.}
{\bf 16} 813
\item {[7]}
Dowker J S and Kirsten K 1999 {\it Class. Quantum Grav.}
{\bf 16} 1917
\item {[8]}
Abouelsaood A, Callan C G, Nappi C R and Yost S A 1987 
{\it Nucl. Phys.} B {\bf 280} 599
\item {[9]}
Callan C G, Lovelace C, Nappi C R and Yost S A 1987
{\it Nucl. Phys.} B {\bf 288} 525
\item {[10]}
Barvinsky A O 1987 {\it Phys. Lett.} {\bf 195B} 344
\item {[11]}
Esposito G, Kamenshchik A Yu, Mishakov I V and Pollifrone G 1995
{\it Phys. Rev.} D {\bf 52} 3457
\item {[12]}
Avramidi I G, Esposito G and Kamenshchik A Yu 1996
{\it Class. Quantum Grav.} {\bf 13} 2361
\item {[13]}
Esposito G, Kamenshchik A Yu and Pollifrone G 1997
{\it Euclidean Quantum Gravity on Manifolds with Boundary}
({\it Fundamental Theories of Physics 85})
(Dordrecht: Kluwer)
\item {[14]}
Moss I G and Silva P J 1997 {\it Phys. Rev.} D {\bf 55} 1072
\item {[15]}
Avramidi I G and Esposito G 1998 {\it Class. Quantum Grav.}
{\bf 15} 1141
\item {[16]}
Gilkey P B 1995 {\it Invariance Theory, the Heat Equation and the
Atiyah--Singer Index Theorem} (Boca Raton, FL: Chemical 
Rubber Company)
\item {[17]}
Esposito G 1999 {\it Class. Quantum Grav.} {\bf 16} 1113
\item {[18]}
Grubb G 1996 {\it Functional Calculus of Pseudodifferential
Boundary Problems} ({\it Progress of Mathematics 65})
(Boston: Birkh\"{a}user)
\item {[19]}
Schr\"{o}der M 1989 {\it Rep. Math. Phys.} {\bf 27} 259
\item {[20]}
Marachevsky V N and Vassilevich D V 1996 
{\it Class. Quantum Grav.} {\bf 13} 645
\item {[21]}
Esposito G 1998 {\it Dirac Operators and Spectral Geometry}
({\it Cambridge Lecture Notes in Physics} vol 12)
(Cambridge: Cambridge University Press)
\item {[22]}
DeWitt B S in {\it Relativity, Groups and Topology II} eds 
B S DeWitt and R Stora (Amsterdam: North--Holland)
\item {[23]}
Eastwood M and Singer I M 1985 {\it Phys. Lett.} A 
{\bf 107} 73
\item {[24]}
Esposito G 1997 {\it Phys. Rev.} D {\bf 56} 2442
\item {[25]}
Seeley R T 1969 {\it Topics in pseudo-differential operators,
C.I.M.E.}, in Conf. on Pseudo-Differential Operators, ed.
L. Nirenberg (Roma: Edizioni Cremonese)
\item {[26]}
Gibbons G W and Hawking S W 1993 {\it Euclidean Quantum Gravity}
(Singapore: World Scientific)
\item {[27]}
Kree P {\it Les noyaux des operateurs pseudo-differentials},
Publications du s\'eminaire de mathematiques de l'Universit\'e
de Bari
\item {[28]}
Hartle J B and Hawking S W 1983 {\it Phys. Rev.} D {\bf 28} 2960

\bye